\documentclass[showpacs,twocolumn,prd]{revtex4}
\usepackage{graphicx}
\usepackage{latexsym}
\usepackage{amsmath}

\newcommand{\beq}{\begin{equation}}
\newcommand{\eeq}{\end{equation}}
\newcommand{\bea}{\begin{eqnarray}}
\newcommand{\eea}{\end{eqnarray}}

\begin{document}

\title{The structure of the strange sea of the proton}


\author{C. \'Avila}
\email{cavila@uniandes.edu.co}
\author{L. Salazar-Garcia}
\author{J.C. Sanabria}
\email{jsanabri@uniandes.edu.co}
\affiliation{Departamento de F\'{\i}sica, Universidad de los Andes, A.A.
4976, Bogota, D.C., Colombia}

\author{J. Magnin}
\email{jmagnin@cbpf.br}
\affiliation{Centro Brasileiro de Pesquisas F\'{\i}sicas, Rua Dr. Xavier
Sigaud 150, Urca 22290-180, Rio de Janeiro, Brazil}

\begin{abstract}
In this work we study the strange sea of the proton using a version
of the Meson Cloud Model containing both, efective and perturbative degrees of
freedom. We construct the $s$ and $\bar{s}$ parton
distributions functions at the initial energy scale, $Q_0^2$, where QCD
evolution starts. The initial $s$ and $\bar s$ pdfs depend on a number
of parameters which we fix by comparison to parameterizations of
the strange sea of the nucleon obtained in a recent global fit to
experimental data, allowing for a $s-\bar{s}$ asymmetry. We show that the
model describes well the strange sea of the proton and argue that it
can be a phenomenologically motivated 
alternative to the usual input parameterizations used in fits to experimental
DIS data. 
\end{abstract}

\maketitle

\section{Introduction}

The composite nature of hadrons in term of quark and gluon degrees of freedom is
today firmly stablished. It is also firmly believed that the internal dynamics
of hadrons is determined by the strong interactions between quarks and gluons,
as governed by Quantum Chromodynamics (QCD). However, a detailed theoretical
description of the hadron structure is still missing because QCD can only be
solved in the perturbative regime, corresponding to the short distance domain
probed in hard collisions, whereas the long distance part of the interaction
requires a non-perturbative treatment usualy supplied by effective models,
latice simulations, etc. It is worth noting that is just the long distance
part of the interaction which is responsible for the hadron as a bound state of
quarks and gluons. Indeed, hadrons are made up of a fixed number of valence quarks
plus a varying number of sea quark-antiquark pairs and gluons, being these last
the ``glue'' which keeps valence quarks together forming a hadron. As a matter
of fact, the sea $q-\bar q$ pairs and gluons which bring together the valence
quarks into a bound state are the so-called {\em intrinsic} sea~\cite{brodsky},
which has to be distinguished from the {\em extrinsic} sea generated by QCD
evolution and consequently, dependent on the energy scale $Q^2$.

From an experimental point of view, the structure of nucleons is by far the best
known hadron structure, existing today a variety of parton distribution function
(pdf) parameterizations extracted from data. Although the analysis of the
existing data has confirmed the impresive success of the Quark Parton Model and
the Dokshitzer-Grivov-Lipatov-Altarelli-Parisi (DGLAP) evolution
equations~\cite{dglap} when describing the nucleon structure, in the Bjorken
regime, in terms of pdf's given as a function of $x$, the fraction of the
nucleon momentum carried by partons, and $Q^2$, the energy scale which in Deep
Inelastic Scattering (DIS) experiments is identified with the 4-momentum
transfer between the lepton and the nucleon, several questions remain still
unsolved. Among them, what is the functional form - and why - of the valence and
sea quark and gluon pdf's at the energy scale $Q_0^2$ where QCD evolution
starts; what is the value of the initial $Q_0^2$ energy scale; what is the 
structure of the so-called intrinsic sea of $q-\bar q$ pairs and gluons; etc. Of
course, finding answers to these questions requires to deal with the long
distance - confining - realm of QCD, thus, as long as there is no solution of
QCD in the low $Q^2$ regime, we have to rely on effective models.

Furthermore, the structure of the nucleon's intrinsic sea of quarks and gluons
reflects the dynamics of non-perturbative QCD. In fact, the form of the initial
pdfs at the $Q_0^2$ initial scale, as well as the relative abundance of the
different quark flavors are the footprint of subtle non-perturbative QCD
dynamical effects. Notice, for instance, that the experimentally observed
$\bar{d}/\bar{u}$ asymmetry~\cite{e866} and the violation of the Gottfried sum rule
(GSR)~\cite{gsr} cannot be described in terms only of perturbative QCD and gluon
splitting~\cite{gsr-models}, as the ratio $\bar{d}/\bar{u}$ would became equal to
one, due to the equal probability of gluon splitting into $d\bar{d}$ or
$u\bar{u}$ pairs. The distribution of strange and anti-strange quarks in the
nucleon sea is another nontrivial aspect of the nucleon structure. From the
experimental side, some evidence has been found on a possible $s-\bar{s}$
asymmetry coming from global fits to data~\cite{global}. Notice that there is no
fundamental symmetry preventing $s(x)\neq \bar{s}(x)$ in the nucleon, provided
that $\int_0^1\left[s(x) - \bar{s}(x)\right] = 0$. On the theoretical side,
speculations about a possible $\left|KH\right>$ component in the nucleon wave
function, where $K,~H$ are virtual Kaon and Hyperon states, leading to a $s-\bar
s$ asymmetry, date since 1987, with the pionering work by Signal and
Thomas~\cite{signal}. Since then on, several models have been proposed in the
literature~\cite{magnin,other}, with different predictions. It is worth to note
also that a $s-\bar{s}$ asymmetry in the nucleon is generated perturbatively
starting at Next-to-Next to Leading Order (NNLO)~\cite{daniel} because at this
order the splitting functions $P_{qq}$ and $P_{q\bar q}$ are different. That
asymmetry, although very small, should compete with the $s-\bar s$ asymmetry
generated by the non-perturbative dynamics of the bound state. 

In this paper we shall investigate to what extent a non-perturbatively generated
$s-\bar s$ asymmetry can describe the strange sea asymmetry found in recent
global fits to experimental data. In order to do that, in Section~\ref{model} we
will revise a model for the generation of the intrinsic $s/\bar s$ sea of the
nucleon, following in Section~\ref{fit} with a comparison of the model with
experimental data and recent parameterizations of the strange sea of the
proton. Section~\ref{conclusions} will be devoted to further discussion and
conclusions.

\section{The non-perturbative strange sea of the proton}
\label{model}

We start by considering a simple picture of the nucleon in the 
infinite momentum frame as being formed by three dressed valence 
quarks - {\it valons}, $v(x)$ - which carry all of its momentum~\cite{hwa}, 
\begin{eqnarray}
v_u(x) &=& \frac{2}{\beta(a_u+1,b_u+1) }x^{a_u}(1-x)^{b_u}\,,\nonumber \\
v_d(x) &=& \frac{1}{\beta(a_d+1,b_d+1) }x^{a_d}(1-x)^{b_d}\,,
\label{eq1}
\end{eqnarray}
with $x$ the fraction of momentum carried by the valon with respect to the
proton. 

\begin{figure}[t]
  \includegraphics[angle=0, height=.325\textheight]{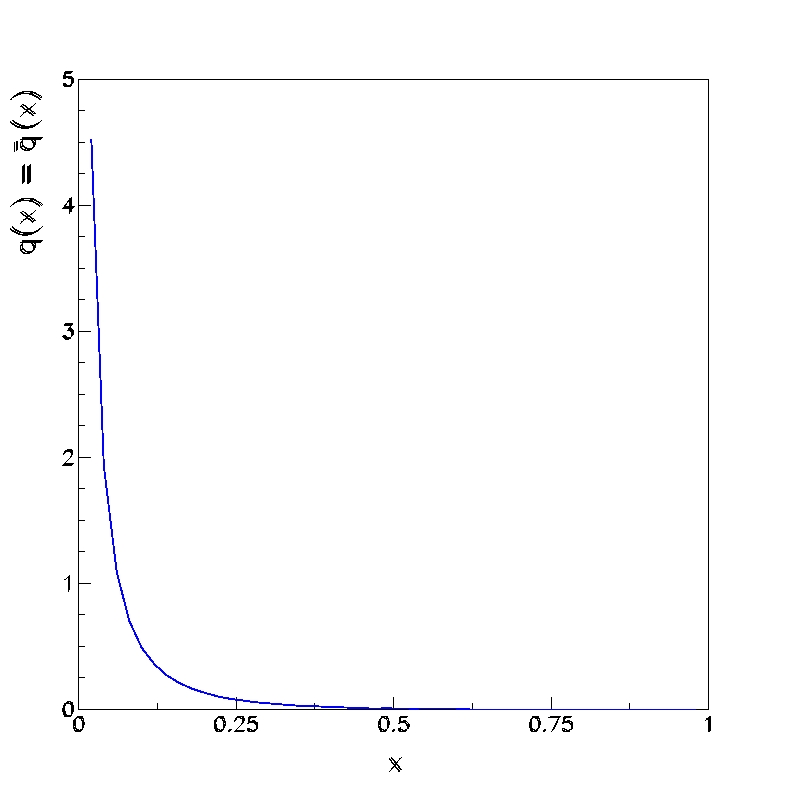}
  \caption{Sea $q/\bar{q}$ parton distribution in the proton as given
    by Eq.~\ref{eq2}.} 
\label{sea}
\end{figure}

In the framework of the Meson Cloud Model (MCM), the nucleon can fluctuate to a  
meson-baryon bound state carrying zero net strangeness. As a first step
in such a process, we may consider that each valon can emit a gluon which,
before interacting, decays perturbatively into a $s\bar s$ pair. The probability
of having such a perturbative $q\bar{q}$ pair can be computed in terms of
Altarelli-Parisi splitting functions~\cite{dglap}  
\bea
P_{gq} (z) &=& \frac{4}{3} \frac{1+(1-z)^2}{z}, \nonumber \\
P_{qg} (z) &=& \frac{1}{2} \left( z^2 + (1-z)^2 \right).
\label{eq2}
\eea
These functions have the physical interpretation as the probability 
of gluon emision and $q\bar{q}$ creation with momentum fraction $z$ 
from a parent quark or gluon respectively. Hence, 
\bea
q(x,Q_0^2) =&& \bar{q}(x,Q_0^2) = N \frac{\alpha_{st}^2(Q_0^2)}{(2\pi)^2}
\times \nonumber \\
&&\int_x^1 {\frac{dy}{y} P_{qg}\left(\frac{x}{y}\right) 
\int_y^1{\frac{dz}{z} P_{gq}\left(\frac{y}{z}\right) v(z)}}
\label{eq3}
\eea
is the joint probability density of obtaining a quark or anti-quark 
coming from  subsequent decays $v \rightarrow v + g$ 
and $g \rightarrow q + \bar{q}$ at some fixed low $Q_0^2$. In eq. (\ref{eq3}),
$N$ is a normalization constant which should scale with the masses of the flavors
being created so that to a heavier flavor, corresponds a smaller $N$.  
Since the valon distribution does not depend on $Q_0^2$ \cite{hwa}, 
the scale dependence in eq.~(\ref{eq3}) only exhibits through the 
strong coupling constant $\alpha_{st}$. The range of values of $Q_0^2$ 
at which the process of virtual pair creation occurs in this approach
is typically below 1 GeV$^2$. A tipical sea quark distribution obtained from
eq. (\ref{eq3}), is shown in Fig. (\ref{sea}), where, in the spirit of keeping
as simple as possible the description of the process , whe have assumed
$v_u=v_d$ and used $a_u=a_d=0.5$, $b_u=b_d=2$ in eqs~(\ref{eq1}).

Once a $s\bar{s}$ pair is produced, it can rearrange itself with the 
remaining valons so as to form a most energetically favored 
Kaon-Hyperon bound state. In order to obtain the Kaon and Hyperon probability
densities in the $\left|KH\right>$ component of the proton wave function, the
well known approach of the recombination model~\cite{das-hwa} has been
used~\cite{magnin}. Notice however that the Kaon and Hyperon probability
densities obtained in this way can be represented in terms of the simple
forms~\cite{magnin} 
\begin{eqnarray}
P_K(x) &=& \frac{1}{\beta(a_K+1,b_K+1) }x^{a_K}(1-x)^{b_K}\,,\nonumber \\
P_H(x) &=& \frac{1}{\beta(a_H+1,b_H+1) }x^{a_H}(1-x)^{b_H}\,,
\label{eq3b}
\end{eqnarray}
which are both properly normalized to unity, as it should be for a bound state
of one Kaon and one Hyperon and in order to cope with the zero net strangeness
of the proton. These forms are also consistent with the valon model for a hadron
made of two partons bound state. The coefficients $a_K,~b_K,~a_H,~b_H$ in
eqs. (\ref{eq3b}) are not independent. In fact, as the Kaon and Hyperon have to
exhaust the momentum of the proton, then
\beq
\int_0^1 {dx \left[xP_H(x) + xP_K(x) \right]}=1\;,
\label{eq4}
\eeq
giving the constraint
\bea
\label{sum}
&&\frac{\Gamma(a_{\mbox{\tiny K}}+b_{\mbox{\tiny K}}+2)\Gamma(a_{\mbox{\tiny K}}+2)}
{\Gamma(a_{\mbox{\tiny K}}+1)\Gamma(a_{\mbox{\tiny K}}+b_{\mbox{\tiny
      K}}+3)} + \nonumber \\
&&\frac{\Gamma(a_{\mbox{\tiny H}}+b_{\mbox{\tiny H}}+2)\Gamma(a_{\mbox{\tiny H}}+2)}
{\Gamma(a_{\mbox{\tiny H}}+1)\Gamma(a_{\mbox{\tiny H}}+b_{\mbox{\tiny
      H}}+3)} = 1\;.
\eea 

Finally, the non-perturbative strange and anti-strange sea distributions in the
nucleon can be computed by means of the two-level convolution formulas 
\bea
\label{eq8a}
s^{NP}(x) &=& N\int^1_x {\frac{dy}{y} P_H(y)\ s_{H}(x/y)},  \\
\bar{s}^{NP}(x) &=& N\int^1_x {\frac{dy}{y} P_K(y)\ \bar{s}_{K}(x/y)},
\label{eq8}
\eea
where the sources $s_{H}(x)$ and $\bar{s}_{K}(x)$ are the probability densities
of the strange valence quark and anti-quark in the Hyperon and Kaon
respectively, evaluated at the hadronic scale $Q_0^2$ \cite{signal}. 
In principle, to obtain the non-perturbative distributions given by 
eqs.~(\ref{eq8}), one should sum over all the strange Kaon-Hyperon 
fluctuations of the nucleon but, since such hadronic Fock states are 
necessarilly off-shell, the most likely configurations are those closest to the
nucleon energy-shell, namely $\Lambda^0K^+$, $\Sigma^+K^0$ and $\Sigma^0K^+$,
for a proton state. For the sake of simplicity, we will only consider a generic
Kaon and Hyperon inside the proton. 

For the $s_{H}(x)$ and $\bar{s}_{K}(x)$ probability densities in eqs. (\ref{eq8a})
and (\ref{eq8}) we also used the simple forms
\begin{eqnarray}
s_H(x) &=& \frac{1}{\beta(a_{sH}+1,b_{sH}+1) }x^{a_{sH}}(1-x)^{b_{sH}}\,,\nonumber \\
\bar{s}_K(x) &=& \frac{1}{\beta(a_{sK}+1,b_{sK}+1) }x^{a_{sK}}(1-x)^{b_{sK}}\,,
\label{eq9}
\end{eqnarray}
according to the valon model. The coefficients $a_{sK},~b_{sK},~a_{sH},~b_{sH}$
in eqs. (\ref{eq9}) have to be determined by comparison to experimental data.

\begin{table}
\caption{\label{table1}Coefficients for $s(x)$ and $\bar{s}(x)$ obtained from
  simultaneous fits to the strange parton distributions funcions in the proton
  as given in Ref.~\cite{global}. $a_H=2.889$ is fixed by the requirement of
  momentum conservation given by eq.~(\ref{sum}).}
\begin{ruledtabular}
\begin{tabular}{ccc}
$b_H$ & $1.563$ & $\pm 0.166$ \\
$a_K$ & $2.403$ & $\pm 0.127$ \\
$b_K$ & $4.164$ & $\pm 0.188$ \\
$a_{sH}$ & $0.669$ & $\pm 0.052$ \\
$b_{sH}$ & $8.539\times 10^{-5}$ & $\pm0.087$ \\
$a_{sK}$ & $23.203$ & $\pm0.085$ \\
$b_{sK}$ & $0.602$ & $\pm 0.002$ \\
$N$ & $0.019$ & $0.001$
\end{tabular}
\end{ruledtabular}
\end{table}

\section{comparison with data}
\label{fit}

\begin{figure}
\includegraphics[angle=0, height=.325\textheight]{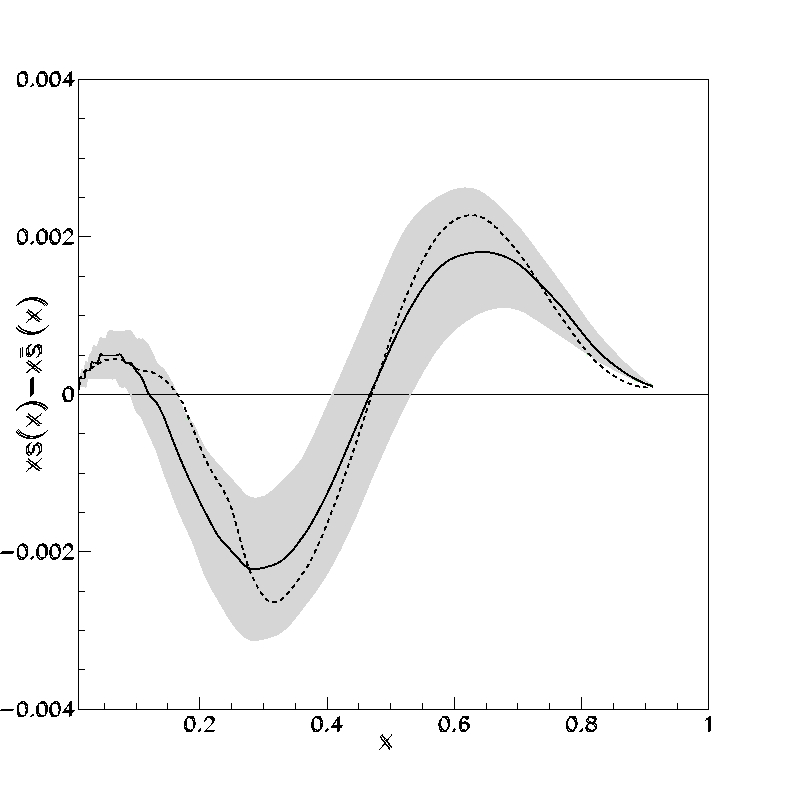}
\caption{\label{fig2}$xs(x)-x\bar{s}(x)$ at $Q^2=20$ GeV$^2$. Dashed line: the
 asymmetry obtained in Ref.~\cite{global}. Full line: the result
  of our fit. The shadow area is the uncertainty of the fit.}
\end{figure}

\begin{figure}
\includegraphics[angle=0, height=.325\textheight]{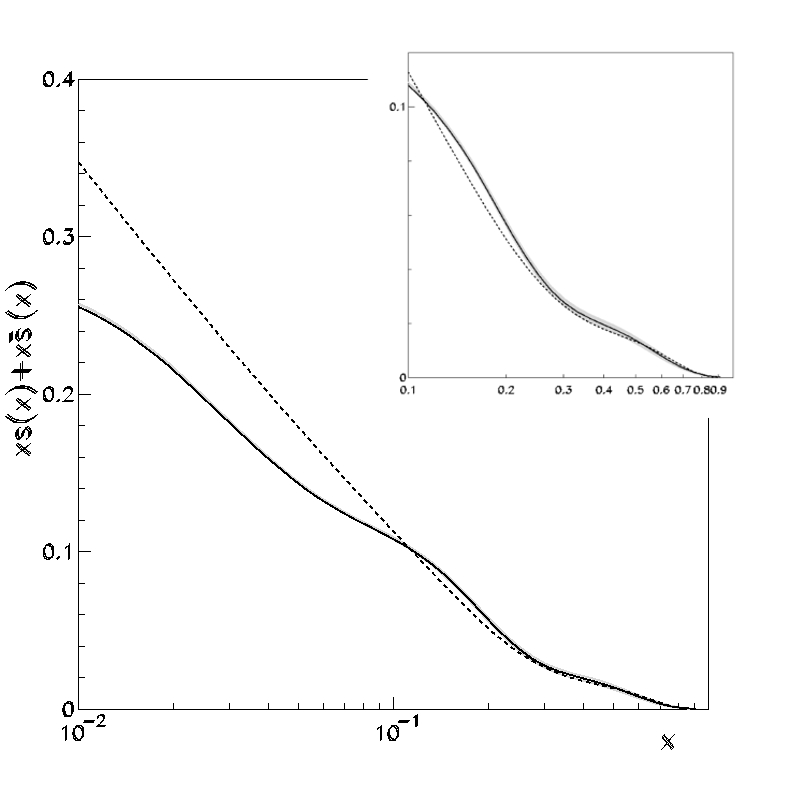}
\caption{\label{fig3}$xs(x)+x\bar{s}(x)$ at $Q^2=20$ GeV$^2$. Dashed line: the
 distribution obtained in Ref.~\cite{global}. Full line: the result of
  our fit. In the insert is shown the result of our fit for $0.1<x<1$. The
  uncertainty of the fit is represented by the shaded area.}
\end{figure}

In order to fix the coefficients of the model, we compare with the $s$ and
$\bar{s}$ parton distribution functions\footnote{Notice however that columns E
  and D in Table 2 in JHEP of Ref.~\cite{global} have to be exchanged - private
  communication with the authors} given in Ref.~\cite{global},
which have been recently obtained in a global fit to DIS data. The comparison
has been done by means of a simultaneous least square fit of the model to
$xs(x)+x\bar{s}(x)$ and $xs(x)-x\bar{s}(x)$ at $Q^2=20$ GeV$^2$. The results of
the fit are shown in Figs.~(\ref{fig2}) and (\ref{fig3}) and in Table~\ref{table1}. 

The procedure of the fit was as follows: {\it i)} a set of initial
values for the parameters was chosen and the non-perturbative $s^{NP}$ and
$\bar{s}^{NP}$ strange quark distribution functions were calculated according to
eqs.~(\ref{eq8a}) and (\ref{eq8}). Then {\it ii)} the $s^{NP}$ and
$\bar{s}^{NP}$ distributions were evolved from $Q_0^2$ up to $Q^2 = 20$ GeV$^2$
and the squared distance
\begin{eqnarray}
\mathcal{S}^2 &=& \sum_{i=1}^{n}\left[\frac{\left(y_{d}(x_i) -
    y_{d}^{th}(x_i)\right)^2}{\sigma_{d}^2(x_i)}\right. \nonumber \\
&& \left. + \frac{\left(y_{s}(x_i) -
    y_{s}^{th}(x_i)\right)^2}{\sigma_{s}^2(x_i)}\right] 
\label{squared}
\end{eqnarray}
was calculated. In eq.~(\ref{squared}), $y_k(x_i) = xs^{NP}(Q^2,x_i) \pm
x\bar{s}^{NP}(Q^2,x_i)$ and $y_k^{th} = xs(Q^2,x_i) \pm x\bar{s}(Q^2,x_i)$, with
index $k=d,s$ refering to the difference and the sum respectively. An arbitrary
error $\sigma_{d/s}^2(x_i)$ corresponding respectively to $10$\% of the value of
the difference and the sum at $x_i$ has been considered to perform the
fit. Finally a new set of parameters was chosen and the procedure has been
repeated until a minimum in $\mathcal{S}^2$ was reached. The package MINUIT has
been used to perform the fit. Notice also that the QCD evolution of the
combination $s+\bar s$ depends on the whole set of parton distribution functions
in the proton, for which we used the pdf's of Ref.~\cite{global}.

\begin{figure}
\includegraphics[angle=0, height=.325\textheight]{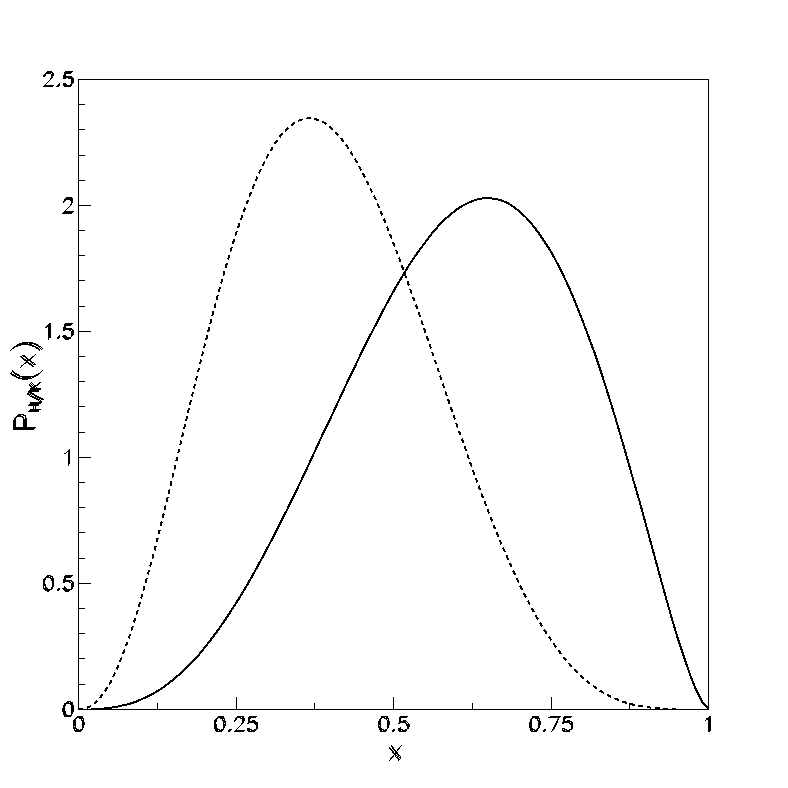}
\caption{\label{fig4} Full line: Hyperon probability density in the
  $\left|KH\right>$ Fock state of the proton. Dashed line: Kaon probability density
in the $\left|KH\right>$.}
\end{figure}

The best fit has been obtained using $Q_0=0.3$ GeV as the starting point for QCD
evolution. As can be seen in Figs.~(\ref{fig2}) and (\ref{fig3}), the model
represents fairly well the strange parton distributions found in
Ref.~\cite{global} for $x \gtrsim 0.1$, whereas for smaller values of $x$ our
results are below the results of the global fit of Ref.~\cite{global}. This can
be due to a deficit in the content of gluons, as seems to be indicated by the fact that a
good fit is only obtained for extremely low values of $Q_0$.

Concerning the model itself, the Kaon and Hyperon distributions functions in the
$\left|KH\right>$ Fock state of the proton are displayed in
Fig.~(\ref{fig4}). The momentum carried by the Hyperon in the $\left|KH\right>$
Fock state is $xP_H(x)=0.6$ while for the Kaon we obtained $xP_K(x)=0.4$,
agreeing with the common intuition that the Hyperon carries more momentum than
the Kaon in the $\left|KH\right>$ component of the proton wave-function.

\begin{figure}
\includegraphics[angle=0, height=.325\textheight]{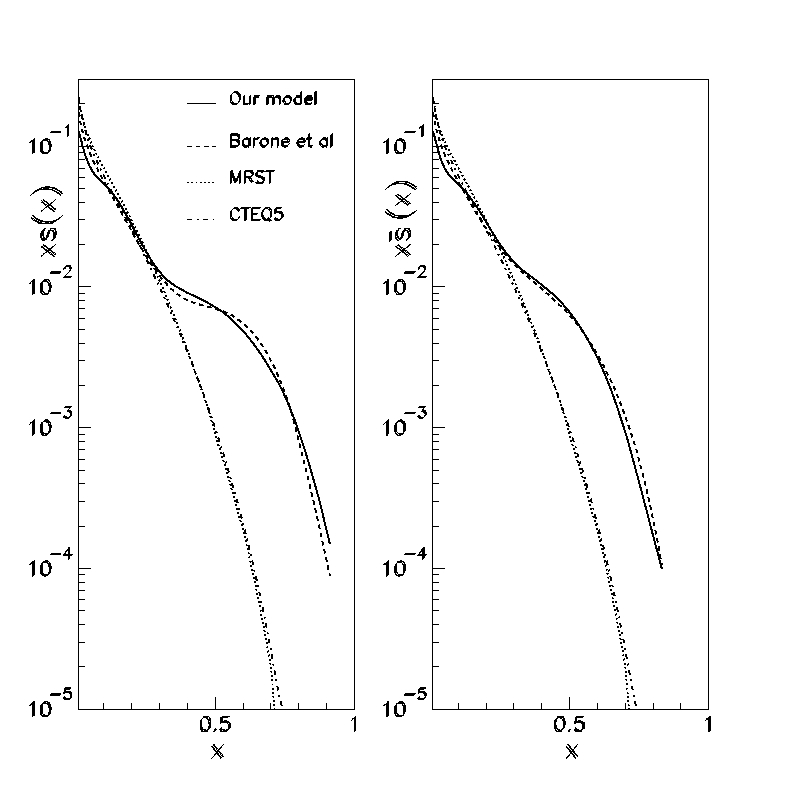}
\caption{\label{fig5} Strange (left) and anti-strange (right) quark
  distributions in the proton at $Q^2 = 20$ GeV$^2$. Our model compared to the 
V. Barone {\it et al.}~\cite{global}, MRST~\cite{mrst} and CTEQ5~\cite{cteq}
strange parton distribution functions.}
\end{figure}

In Fig.~(\ref{fig5}) the strange and anti-strange quark distributions at $Q^2 =
20$ GeV$^2$ are shown and compared to the $xs$ and $x\bar{s}$ distributions
found in Ref.~\cite{global}. We also compare to the MRST~\cite{mrst} and
CTEQ5~\cite{cteq} strange quark pdfs, which have been determined imposing
$s=\bar{s}$. As shown in the figure, while our strange quark and anti-quark pdfs
and those of Ref.~\cite{global} are consistent in the full range $0.01 < x < 1$,
they deviate from the behavior shown by the MRST and CTEQ5 sets at $x\gtrsim 0.3$.

\section{Conclusions}
\label{conclusions}

In this paper we have presented a model for the non-perturbative structure of
the strange sea of the proton. In the model, the non-perturbative $s^{NP}$ and
$\bar{s}^{NP}$ intrinsic sea quark distributions of the proton are given in
terms of a convolution of Hyperon and Kaon probability densities and valence
quark distributions inside a $\left|KH\right>$ component of the proton wave
function. This naturally generates an asymmetry in the momentum distributions of
the strange and anti-strange quarks in the proton, since the Hyperon, being
heaviest than the Kaon, carries more momentum in the $\left|KH\right>$ wave
funtion component. The model depends on eight parameters which
have to be fixed by fits to experimental data. 

Parameters of the model have been fixed by fits to the strange quark
distributions found in a recent global fit to DIS data~\cite{global}. As shown
in section \ref{fit}, the model describes qualitatively well the behavior of the
$s-\bar{s}$ as well as $s+\bar{s}$ distributions, being this agreement better in
the region $x\gtrsim 0.1$. For lower $x$, the $s+\bar s$ distribution is below
the corresponding curve given by the distributions of Ref.~\cite{global}. The
probability of the $\left|KH\right>$ fluctuation of the proton is about
$0.02$\%, as given by the parameter $N$.

The fact that the model does not describe well the $s+\bar s$ distribution at
$x < 0.1$ is expected by several reasons. First of all, the parton distributions
found in Ref.~\cite{global} were determined in a global fit to DIS data,
instead our $s$ and $\bar s$ distributions have been determined by fits to the
$s-\bar s$ and $s + \bar s$ distributions found in ~\cite{global}. This
restrict the space of parameters allowed to the fit. Second, the gluon
distribution found in Ref.~\cite{global} seems to be insufficient to generate
enough $s/\bar{s}$ quarks at low momentum, as evidenced by the fact that while
the $s-\bar s$ distribution is well described in the whole range $0 < x < 1$,
the $s+\bar s$ is not. This could also explain why a good fit is
obtained only for extremely low values of $Q_0$, the scale at which perturbative
QCD evolution starts. And third, a meaningful comparison of the model has to
be done through a global fit to experimental data.

Notice also that no NNLO effects have been included in the QCD evolution of the
strange parton distributions. The inclusion of those effects should produce a
slightly bigger non-perturbative contribution to the proton wave funtion to
compensate the opposite asymmetry which arises at NNLO~\cite{daniel}.

Finally we would like to emphasize that the model presented here can be taken
as a phenomenologically motivated alternative for the description of the input
strange and anti-strange quark distributions used in fits to experimental
data. Notice that the model allows for a full representation of the non-perturbative
processes inside the proton in terms of well known mechanisms such as the
splitting of quarks and gluons and recombination, which after all must account
for the dynamics of the proton as a bound state. Another interesting aspect of
the model is that it can shed light on the parton structure of real Kaons and
Hyperons, since the structure of the strange sea of the proton is related to the
strange valence quark distributions at a low $Q^2$ scale of the former strange
mesons and baryons.

\section*{Acknowledgements}
Support for this work has been received from the ``Fundaci\'on para la
promoci\'on de la Investigaci\'on y la tecnologia'', Banco de la Rep\'ublica de
Colombia under contract No. 2407 and FAPERJ, under project
No. E-26/110.266/2009. J.M. acknowledges the warm hospitality in the Physics
Department, Universidad de los Andes, where part of this work has been done.

\end{document}